\RequirePackage{fix-cm,amssymb,pdfsync}
\documentclass[twocolumn]{svjour3}          
\smartqed  
\usepackage{graphicx}
\begin{document}

\title{Superconductivity in striped and multi-Fermi-surface Hubbard
models: From the cuprates to the pnictides
}


\author{Thomas A. Maier         
}


\institute{Thomas A. Maier \at
              Computer Science and Mathematics Division and Center for Nanophase Materials Sciences \\
			Oak Ridge National Laboratory
              Tel.: +865-539-6868\\
              \email{maierta@ornl.gov}           
}

\date{Received: date / Accepted: date}

\maketitle

\begin{abstract}

  Single- and multi-band Hubbard models have been found to describe many 
  of the complex phenomena that are observed in the cuprate and 
  iron-based high-temperature superconductors. Simulations of these 
  models therefore provide an ideal framework to study and understand 
  the superconducting properties of these systems and the mechanisms 
  responsible for them. Here we review recent dynamic cluster quantum 
  Monte Carlo simulations of these models, which provide an unbiased 
  view of the leading correlations in the system. In particular, we 
  discuss what these simulations tell us about superconductivity in the 
  homogeneous 2D single-orbital Hubbard model, and how charge stripes 
  affect this behavior. We then describe recent simulations of a bilayer 
  Hubbard model, which provides a simple model to study the type and 
  nature of pairing in systems with multiple Fermi surfaces such as the 
  iron-based superconductors.

\end{abstract}

Phenomenologically, the cuprate and iron-based high-temperature
superconductors share in common that superconductivity occurs by doping
into an antiferromagnetic Mott or spin density wave state. The magnetism
in these compounds originates from partially filled Cu or Fe d-orbitals
which energetically lay at the Fermi energy. A reasonable starting point
for a theoretical description of these compounds is therefore given by a
Hubbard model which describes moment formation due to a strong local
Coulomb repulsion between the electrons on the d-orbitals. In the case
of the cuprates, a single electronic band mainly of Cu
d$_{x^2-y^2}$-character crosses the Fermi surface and a single-band
Hubbard model has therefore been argued to provide a simple framework to
describe the low-energy physics of this class of materials
\cite{anderson_resonating_1987,zhang_effective_1988}. In the iron-based
materials, one has multiple Fermi surfaces formed by Bloch states
originating from several of the iron d-orbitals. Therefore, one needs to
start with a multi-orbital Hubbard model to describe the more complex
electronic structure of these systems
\cite{graser_near-degeneracy_2009,kuroki_unconventional_2008}. A
two-band model has also been used to study phase separation in cuprate
superconductors \cite{PhysRevB.78.165124}.

Here, we review recent dynamic cluster quantum Monte Carlo calculations 
of these models 
\cite{maier_structure_2006,maier_pairing_2006,maier_dynamic_2010,maier_bilayer}.  
In particular, we will focus on what these simulations tell us about the 
nature of pairing and the effect of charge stripes in the single-band 2D 
Hubbard model, as well as the leading pairing correlations in a bilayer 
Hubbard model with multiple Fermi surfaces. 

The Hamiltonian for the 2D Hubbard model we will study can be written as 
\begin{equation}
  {\cal H}=-t\sum_{\langle ij\rangle,\sigma} 
  (c^\dagger_{i\sigma}c^{\phantom\dagger}_{j\sigma} +h.c.) + U\sum_i 
  n_{i\uparrow}n_{i\downarrow}\,. \label{eq:1} \end{equation}
Here $t$ is a nearest neighbor hopping parameter, $U$ is an on-site 
Coulomb repulsion and $\langle ij\rangle$ implies summation over nearest 
neighbor pairs only. In the following we will measure energies in units 
of $t$. The single band with dispersion $\varepsilon_k=-2t(\cos k_x+\cos 
k_y)$ describes the electronic states near the Fermi energy in the 
cuprate materials. 

As discussed in the introduction, a realistic description of the 
pnictides requires a multi-orbital Hubbard model describing the 5 iron 
d-orbitals near the Fermi energy and their intra- and inter-orbital 
Coulomb, Hund's rule and pair hopping interactions. Such a model is too 
complex to simulate with quantum Monte Carlo approaches, and so we will 
instead study a simpler two-orbital model with only intra-orbital 
Coulomb interactions. This model will be realized by a bilayer Hubbard 
model. Its Hamiltonian is given by
\begin{eqnarray}
  H&=&-t\sum_{\langle ij\rangle 
    m\sigma}(c^\dagger_{im\sigma}c^{\phantom\dagger}_{jm\sigma}+{\rm 
    h.c.})- 
  t_\perp\sum_{i\sigma}(c^\dagger_{i1\sigma}c^{\phantom\dagger}_{i2\sigma}+{\rm 
    h.c.})\nonumber\\
  &+&U\sum_{im}n_{im\uparrow}n_{im\downarrow}\,. \label{eq:2} 
\end{eqnarray}
Here the layers are indexed by $m$, each layer is described by the 
Hamiltonian in Eq.~\ref{eq:1} and $t_\perp$ is an additional hopping 
parameter between neighboring sites in the bi-layer model. This model 
provides a simplified two-orbital system in which one can study the type 
of pairing that can occur in systems with multiple Fermi surfaces such 
as the iron-pnictides.

In order to analyze these models, we will use a dynamic cluster quantum 
Monte Carlo approximation\linebreak (DCA/QMC) 
\cite{hettler_nonlocal_1998,jarrell_quantum_2001,maier_quantum_2005}.  
The DCA maps the bulk lattice problem onto an effective periodic cluster 
cluster embedded in a dynamic mean-field that is designed to represent 
the rest of the system.  The effective cluster problem is then solved 
using a quantum Monte Carlo algorithm. The results discussed in this 
paper were obtained with a Hirsch-Fye method
\cite{jarrell_quantum_2001}. DCA/QMC calculations of the 2D Hubbard 
model have found many phenomena that are also observed in the cuprates, 
including an antiferromagnetic Mott state, d-wave superconductivity as 
well as pseudogap behavior \cite{maier_quantum_2005}. It therefore 
provides an interesting framework to study many of the open questions in 
the field. 

Formally, the quantity of interest to study the nature of pairing in 
these models is given by the two-particle irreducible vertex in the 
particle-particle channel, $\Gamma^{pp}_{\rm irr}(k,k')$ 
\cite{maier_structure_2006}.  Here $k=({\bf k},i\omega_n)$ with 
$\omega_n$ a fermion Matsubara frequency and we are interested in the 
singlet pairing channel. This quantity describes the scattering of two 
electrons with momenta ${\bf k}$ and $-{\bf k}$ and anti-parallel spins 
to a state with momenta ${\bf k}'$ and $-{\bf k}'$ and therefore 
describes the pairing interaction.  Together with the single-particle 
Green's function $G(k)$, it enters the Bethe-Salpeter equation 
\begin{equation}
  -\frac{T}{N}\sum_k \Gamma^{pp}_{\rm 
    irr}(k,k')G(k')G(-k')\Phi_\alpha(k') = \lambda_{\alpha} 
  \Phi_\alpha(k) \label{eq:BSE} \end{equation}
which provides information on the strength ($\lambda_\alpha$) and 
momentum and frequency structure ($\Phi_\alpha(k)$) of the leading 
pairing correlations in the system \cite{maier_structure_2006}. At 
$T_c$, $\lambda_\alpha=1$ and $\Phi_\alpha(k)$ becomes identical to the 
superconducting gap. In the 2D Hubbard model, at low temperature, one 
finds that the eigenvector corresponding to the leading eigenvalue has a 
d-wave $\cos k_x-\cos k_y$ momentum dependence.

Previous DCA/QMC simulations of the 2D Hubbard model 
\cite{maier_structure_2006,maier_pairing_2006} have found that the 
momentum and frequency dependence of the pairing interaction 
$\Gamma^{pp}_{\rm irr}(k,k')$ is similar to that of the spin 
susceptibility $\chi(k-k')$, providing evidence that that pairing 
interaction in this model is carried by spin fluctuations.

In a spin fluctuation picture, one can naturally understand the drop of 
$T_c$ with doping on the overdoped side of the cuprate phase diagram, 
since the spin-fluctuations are weakened by doping away from the 
antiferromagnetic parent state. On the other hand, the drop of $T_c$ 
with underdoping is difficult to understand in a picture where pairing 
is mediated by spin fluctuations, since one would expect them to get 
stronger when the system is doped towards the Mott state. To investigate 
this issue we show in Fig.~1a the temperature versus doping 
superconducting phase diagram of the 2D Hubbard model, calculated with 
DCA/QMC on an 8-site cluster with U=8. One sees that these calculations 
correctly predict the experimentally observed dome-like structure of the 
superconducting phase diagram, with $T_c$ dropping with both over- and 
underdoping.    \begin{figure}
  [htbp] \includegraphics[width=3.5in]{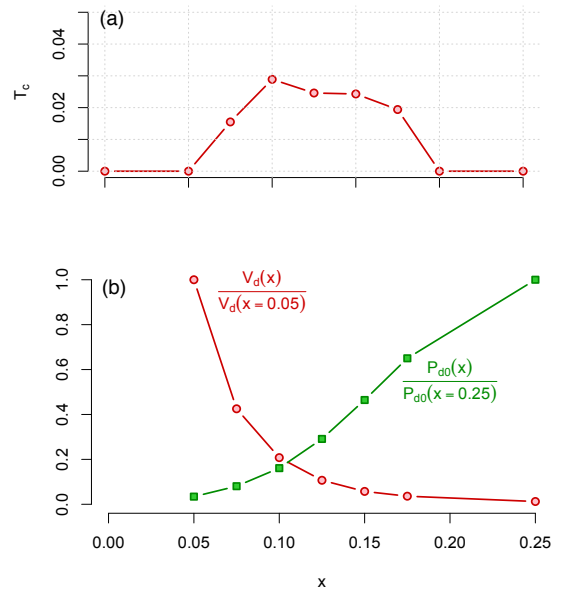} \caption{ (a) 
    Superconducting transition temperature $T_c$ versus doping $x$ for 
    the 2D Hubbard model with $U=8$ calculated with DCA/QMC on an 8-site 
    cluster. (b) Normalized interaction strength $V_d$ and "intrinsic" 
    pair-field susceptibility $P_{d,0}$ versus doping $x$ calculated at 
    a temperature $T=0.125$.  \label{fig:1}} \end{figure}

In order to analyze how this behavior arises from the Bethe-Salpeter 
equation (\ref{eq:BSE}), we have calculated the "intrinsic" pair-field 
susceptibility projected onto the leading eigenvector, $P_{d,0} = T/N 
\sum_k \Phi_d(k)^2 G(k)G(-k)$ and the strength of the pairing 
interaction $V_d$ from $V_d P_{d,0} = \lambda_d$. The doping dependence 
of these quantities calculated at a low temperature above $T_c$ are 
shown in Fig.~1b. As one would expect from a spin-fluctuation based 
pairing interaction, $V_d$ rises monotonically with decreasing doping 
towards the Mott insulator. In contrast, $P_{d,0}$ decreases with 
decreasing doping and goes to zero as one approaches the Mott state. The 
drop of $T_c$ on the underdoped side therefore is caused by the strong 
Mott quasiparticle renormalization when the doping is close to zero.

Despite the strong pairing interaction $V_d$ in the underdoped system, 
$T_c$ is small because of the Mott quasiparticle degradation. It has 
been suggested that in this case, $T_c$ may be enhanced in a striped 
state, in which the system is separated into hole-rich regions with good 
hole mobility, and hole-poor regions with strong antiferromagnetic 
correlations. To explore this prediction, we have performed DCA/QMC 
calculations of a striped 8$\times$4-site cluster 
\cite{maier_dynamic_2010}. The charge stripes were imposed by hand, by 
adding a site-dependent external potential $V_i$ that couples to the 
charge density on site $i$.  $V_i$ is chosen to have a maximum amplitude 
$V_0$ and to vary smoothly along the long 8-site x-direction of the 
cluster while being constant along the 4-site y-direction. The variation 
of $V_i$ along the x-direction is shown at the bottom of Fig.~2a. Here 
we are interested in a striped inhomogeneity with period 8. The 
calculated site-filling is displayed at the top of Fig.~2a and found to 
follow closely the variation of the external potential.  \begin{figure}
  [htbp] \includegraphics[width=3.5in]{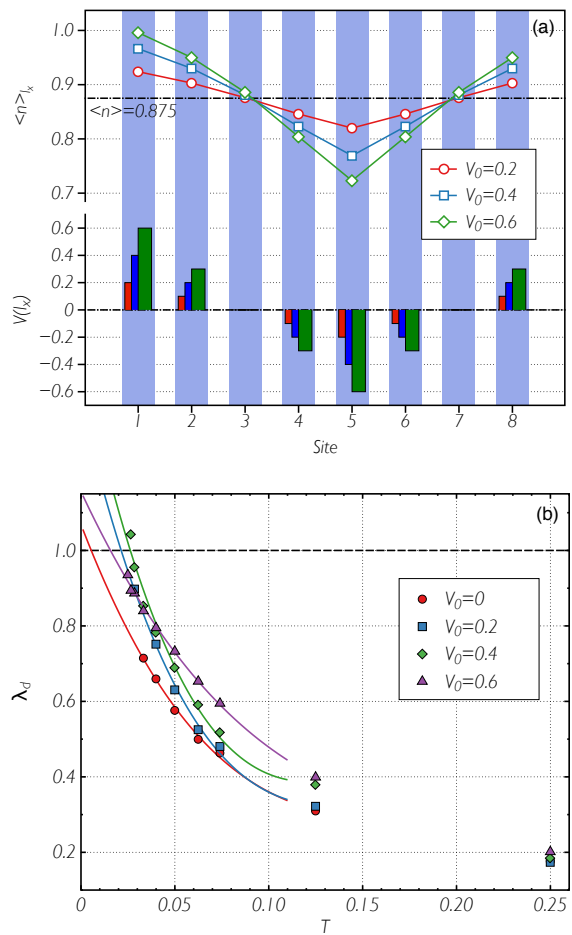} \caption{DCA/QMC 
    simulation of an inhomogeneous 8$\times$4 cluster with a charge 
    stripe. (a) External potential $V(l_x)$ and resulting charge density 
    $\langle n\rangle_{l_x}$ along the long (8-site) direction of the 
    cluster for different magnitudes $V_0$. (b) Leading eigenvalue of 
    the particle-particle Bethe-Salpeter equation (\ref{eq:BSE}) versus 
    temperature for different $V_0$.  \label{fig:1}} \end{figure}

In order to keep the problem computationally tractable, we have averaged 
over different stripe origins along the x-direction, before computing 
the mean-field medium \cite{maier_dynamic_2010}. This corresponds to a 
situation where the stripe order is short-ranged, over the length-scale 
of the cluster, but is translationally invariant on longer macroscopic 
length scales.

The temperature dependence of the leading (d-wave) eigenvalue of the 
Bethe-Salpeter equation (\ref{eq:BSE}) computed for different amplitudes 
$V_0$ of the potential is shown in Fig.~2b. As one can see, the pairing 
correlations are indeed enhanced by the charge stripe and $T_c$, i.e.  
the temperature where $\lambda_d(T)$ crosses 1, is increased.  One also 
sees that there is an optimum inhomogeneity at around $V_0=0.4$ for 
which $T_c$ is maximized. For larger $V_0$, $T_c$ is found to drop 
again.

Finally, we discuss recent DCA/QMC results for the bilayer Hubbard model 
\cite{maier_bilayer}. This model provides a simple system with multiple 
Fermi surfaces, in which one can study the type of pairing that can 
occur in systems such as the iron-pnictides. Due to the inter-layer 
hopping $t_\perp$, one has a bonding and an anti-bonding band split by 
$2t_\perp$ for $U=0$. The bonding and anti-bonding Fermi surfaces of the 
non-interacting system are shown in Fig.~3 (a) and (b) for a filling 
$\langle n\rangle =0.95$ and two values of $t_\perp/t$. As $t_\perp/t$ 
increases the Fermi surfaces shrink and for $\langle n\rangle =1$ the 
non-interacting system becomes a band-insulator for $t_\perp > 4$. Weak 
coupling calculations for the doped system 
\cite{bulut_nodeless_1992,liechtenstein_s-wave_1995} have found a 
$d_{x^2-y^2}$-like gap for $t_\perp/t=0.5$ and a fairly isotropic 
"$s^\pm$" gap that changes sign between the bonding and anti-bonding 
Fermi surfaces for $t_\perp/t=2.0$, as schematically illustrated in 
Fig.~3 (a) and (b).  

\begin{figure}
  [htbp] \includegraphics[width=3.5in]{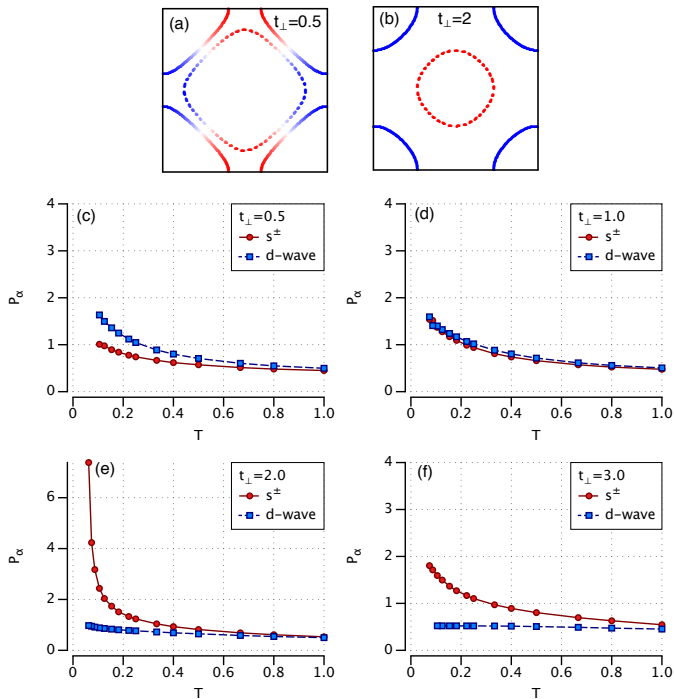} \caption{ DCA/QMC 
    simulation of a bilayer Hubbard model with inter-layer hopping 
    $t_\perp$. The bonding (solid lines) and anti-bonding (dashed lines) 
    Fermi surfaces of the non-interacting system for (a) $t_\perp=0.5$ 
    and (b) $t_\perp=2$ for a filling $\langle n\rangle=0.95$. A 
    $d_{x^2-y^2}$ gap structure is illustrated in for the 
    $t_\perp/t=0.5$ Fermi surface and an $s^\pm$ gap structure is shown 
    for $t_\perp/t=2.0$. (c) The $d$-wave and $s^\pm$ pair-field 
    susceptibilities $P_\alpha$ versus temperature for different values 
    of the inter-layer hopping $t_\perp$. As $t_\perp$ increases, the 
    leading instability changes from $d$-wave to $s^\pm$ and at larger 
    values of $t_\perp$ the superconducting pair-field susceptibility is 
    suppressed. \label{fig:1}} \end{figure}

Here we describe the results of DCA/QMC calculations which provide an 
unbiased treatment of this model and allow us to reach lower 
temperatures than previous finite size quantum Monte Carlo calculations 
\cite{bulut_nodeless_1992,scalettar_magnetic_1994,hetzel_pairing_1994,dos_santos_magnetism_1995,bouadim_magnetic_2008}.  
The superconducting response is studied by calculating the pairfield 
susceptibility \begin{equation}
  P_\alpha(T) = \int_0^\infty d\tau \langle 
  \Delta_\alpha(\tau)\Delta^\dagger_\alpha(0)\rangle\,.  \end{equation}
Here $\Delta^\dagger=1/\sqrt{N}\sum_k g(k) 
c^\dagger_{k\uparrow}c^{\phantom\dagger}_{-k\downarrow}$ and $g(k)=\cos 
k_x-\cos k_y$ for the $d_{x^2-y^2}$-wave and $\cos k_z$ for the $s^\pm$ 
case. In Fig.~3 (c)-(f), results for these pair-field susceptibilities 
versus temperature are shown for different values of $t_\perp$. For 
$t_\perp=0.5$, the d-wave response is larger than the $s^\pm$ 
susceptibility. For $t_\perp=1$, both channels are almost degenerate, 
while for $t_\perp=2$, the $s^\pm$ is the leading response and diverges 
at a relatively high temperature around $0.05t$. For $t_\perp=3$, the 
pair-field response is significantly weakened. This near-degeneracy of 
the $d_{x^2-y^2}$ and $s^\pm$ pair-field correlations for intermediate 
values of $t_\perp/t$ has been similarly observed in 
fluctuation-exchange calculations of realistic 5-orbital model 
calculations of the iron-pnictides 
\cite{graser_near-degeneracy_2009,kuroki_unconventional_2008}.

A further analysis of the pairing interaction in the bilayer model 
\cite{maier_bilayer} shows that the observed $t_\perp$-dependence of the 
pairfield susceptibilities can be understood in terms of the 
$t_\perp$-dependence of the spin-fluctuation spectral weight. For small 
$t_\perp$, the
intra-layer spin fluctuations are dominant and give rise to the 
$d_{x^2-y^2}$ response. As $t_\perp$ increases the dominant 
spin-fluctuations change from intra- to inter-layer, which
give rise to the $s^\pm$ pairing. For large $t_\perp \gtrsim 3$, the 
inter-layer spin fluctuations become gapped and contribute less to the 
pairing.

To conclude, we have reviewed dynamic cluster quantum Monte Carlo
simulations of models of unconventional superconductors, including a 2D
Hubbard model with charge stripes and a bilayer Hubbard model with
multiple Fermi surfaces. We have shown that charge stripes in the 2D
Hubbard model can lead to a significant enhancement of
superconductivity. For the bilayer model, we have found a transition of
the leading pairing instability from a $d_{x^2-y^2}$- wave to an $s^\pm$
state with increasing inter-layer hopping $t_\perp/t$. We have also
discussed how the superconducting behavior of these models can be
understood in terms of a spin-fluctuation picture.

%

\begin{acknowledgements}
We would like to acknowledge useful discussions with D.J. Scalapino, 
T.C. Schulthess, G. Alvarez and M. Summers. This research was conducted 
at the Center for Nanophase Materials Sciences, which is sponsored at 
Oak Ridge National Laboratory by the Office of Basic Energy Sciences, 
U.S. Department of Energy.  This research was enabled by computational 
resources of the Center for Computational Sciences at Oak Ridge National 
Laboratory.
\end{acknowledgements}

\bibliographystyle{spmpsci}      
\bibliography{mybib}   


\end{document}